\documentclass[]{svjour3}  
\usepackage{epsfig} 
\usepackage{graphicx}
\usepackage{dcolumn}
\usepackage{bm}
\usepackage[usenames]{color}
\usepackage{ulem} 
\usepackage{amsmath}
\usepackage{amssymb}
\usepackage[]{natbib}

\journalname{Bulletin of Mathematical Biology}

\begin{document}

\title{Maximal Sensitive Dependence and  the Optimal Path to Epidemic Extinction}

\author{Eric Forgoston \and Simone Bianco \and Leah B. Shaw  \and Ira B. Schwartz}

\institute{Eric Forgoston \at
 Nonlinear Systems Dynamics Section, Plasma Physics Division, Code 6792,
 U.S. Naval Research Laboratory, Washington, DC 20375, USA\\
 \email{eric.forgoston.ctr@nrl.navy.mil}
\and Simone Bianco \at Department of Applied Science, The College of William
\& Mary, P.O. Box 8795, Williamsburg, VA 23187-8795, USA\\
 \email{sbianco@wm.edu}
\and Leah B. Shaw \at Department of Applied Science, The College of William \&
Mary, P.O. Box 8795, Williamsburg, VA 23187-8795, USA\\
\email{lbshaw@wm.edu}
 \and Ira B. Schwartz \at  Nonlinear Systems Dynamics Section, Plasma Physics Division, Code 6792,  U.S. Naval Research Laboratory, Washington, DC 20375, USA\\
 \email{ira.schwartz@nrl.navy.mil}
}

\date{Received: date / Accepted: date}

\maketitle

\begin{abstract}
Extinction of an epidemic or a species is a rare event that occurs due to a
large, rare stochastic fluctuation.  Although the extinction process is dynamically
unstable, 
 it follows an optimal path that maximizes the probability of
extinction.  We show that the optimal path is also directly related to the
finite-time Lyapunov exponents of the underlying dynamical system in that the
optimal path displays maximum sensitivity to initial conditions.  We consider several stochastic epidemic models, and examine the extinction
process in a dynamical systems framework.  Using the
dynamics of the finite-time Lyapunov  exponents as a constructive tool, we
demonstrate  that the dynamical systems
viewpoint of extinction 
evolves naturally toward the optimal path.

\end{abstract}

\section{Introduction}\label{sec:intro}
Control and  eradication of infectious diseases are among the most important
goals  for improving public
health.  Although the global eradication of a disease (e.g. smallpox) has
been achieved~\citep{breari80}, it is difficult to accomplish and remains a public health target
for polio, malaria, and
many other diseases, including childhood diseases~\citep{ayhezaolco00}.  The
global eradication of a disease is not the only type of disease extinction
process.  For example, a disease may ``fade out'' or become locally extinct in
a region.  Since the extinction is local, the disease may be reintroduced from
other regions~\citep{Grassly2005}.  Additionally, extinction may also occur to
individual strains of a multistrain disease~\citep{Minayev2009}, such as
influenza or dengue fever.  It is worth noting that the extinction process is
also of interest in the fields of evolution and ecology.  As an example, bio-diversity arises from the interplay between the introduction and
extinction of species~\citep{Azaele2006,BanMar09}.

In order to  predict the dynamics of disease outbreak and spread as
well as to implement control strategies which promote disease
extinction, one must investigate how model parameters affect the
probability of extinction.  For example,~\citet{dyscla08} showed that disease control and
extinction depend on both epidemiological and sociological parameters
determining epidemic growth rate.  Additionally, since extinction occurs in finite
populations, another factor in determining extinction risk is  the
local community size~\citep{Bartlett1957,Bartlett1960,Keeling1997,Conlan2007}.
Further complications arise from the fact that in general, stochastic effects
cause random transitions within the discrete, finite populations.  These stochastic effects may be intrinsic to the system or may arise from the
external environment.  Small population size, low contact frequency for
frequency-dependent transmission, competition for resources, and evolutionary
pressure~\citep{DeCastro2005}, as well as heterogeneity in populations and
transmission~\citep{Lloyd2007} may all be determining factors for extinction to
occur.  Other factors which can affect the risk of extinction include the
nature and strength of the noise~\citep{Melbourne2008}, disease outbreak amplitude~\citep{Alonso2006}, and seasonal phase
occurrence~\citep{stolhu07}.

In large populations, the intensity of the intrinsic noise is often quite
small.  However, it is possible that a rare, large fluctuation which occurs
with finite probability 
can cause the system to reach the extinct state.
Since the extinct state is absorbing due to effective stochastic forces,
eventual extinction is guaranteed when there is no source of
reintroduction~\citep{bartlett49,Allen2000,gar03}.  However,
because fade outs are usually rare events in large populations, typical time scales for extinction may be
extremely long.

Birth-death systems~\citep{gar03,vanKampen_book}, which possess intrinsic
noise, form an important class of
stochastic processes. These systems have 
been used in the field of mathematical
epidemiology~\citep{bar61,andbri00,chguro07}.  In these systems, the intrinsic
noise arises from the discreteness of the individuals (or species) and the
randomness of their interactions. To  predict probabilities of events occurring at
certain times, a description of the stochastic system is
provided using the master equation formalism.  Stochastic master equations are
commonly used in statistical physics when dealing with chemical reaction
processes~\citep{Kubo63}.

For systems with a large number of individuals, a van Kampen system-size
expansion may be used to approximate the master
equation by a Fokker-Planck equation~\citep{gar03,vanKampen_book}.  However,
the technique fails in determining the probability of large
fluctuations~\citep{gamoto96,elgkam04}.  Instead, in this article, we will
employ an eikonal approximation to recast the problem in terms of an effective
classical Hamiltonian system~\citep{kammee08}.  With high probability, the
intrinsic noise will induce extinction of the disease or species along a
heteroclinic trajectory in the phase space of the classical Hamiltonian flow.
This trajectory is known as the optimal path (to extinction).

It is highly desirable to locate the optimal path since 
the extinction rate depends on the probability to traverse this
path.  Additionally, the
effect of a control strategy on the extinction rate can be determined by its
effect on the optimal path~\citep{dyscla08}.  Through the use of the eikonal
approximation, we consider a mechanistic dynamical systems problem rather than
the original stochastic problem.  Unlike other methods, this approach enables one both to estimate
extinction lifetimes and to draw conclusions about the path taken to extinction.  This more detailed
understanding of how extinction occurs may lead to new stochastic control
strategies~\citep{dyscla08}.

In general, the optimal path to extinction is an unstable dynamical object.  Therefore,
many researchers have investigated the scaling of extinction rates with
respect to a parameter near a bifurcation point, where the dynamics are
slow~\citep{Doering2005,kammee08,Kamenev2008b,dyscla08}.  The analytical
calculation of extinction rates far from a bifurcation point is much more
difficult, and thus, researchers often resort to averaging over many stochastic
runs~(e.g.~\cite{Shaw2009}).  The numerical computation of the optimal
path trajectory has been achieved in the past using a shooting
method~\citep{kammee08}.  However, since the procedure is very sensitive to boundary
conditions,   
it is difficult to implement when analyzing paths far away from bifurcation
points or when analyzing high-dimensional models.

In this article we develop a novel method for computing the optimal extinction
trajectory that relies on the calculation of the dynamical system's finite-time Lyapunov exponents (FTLE).  The classical Lyapunov exponent provides a
quantitative measure of how infinitesimally close particles in a dynamical
system behave asymptotically as time 
$t\to\pm\infty$~\citep{guchol86}.  Similarly,
the FTLE provides a quantitative measure of how much nearby particles separate
after a specific amount of time has elapsed.

The FTLE fields were used by~\citet{pie91} and~\citet{pieyan93} to
characterize structures, including mixing
regions and transport barriers, in the atmosphere.  Later, these structures
were quantified more rigorously by introducing the idea of Lagrangian Coherent
Structures (LCS)~\citep{hall00,hall01,hall02,shlema05,leshma07,brawig09}.  The
definition of a LCS as a ridge of the FTLE field was introduced
by~\citet{hall02} and formalized by~\citet{shlema05}.  The FTLE method
provides a measure of how sensitively the system's
future behavior depends on its current state, and the LCS, or ridge, is a
structure which has locally maximal FTLE value.  In this article, we will show
that the system displays maximum sensitivity near the optimal
extinction trajectory, which enables us to dynamically evolve toward the optimal escape trajectory using FTLE
calculations.

In this article, we consider three standard epidemic models that contain intrinsic or external
noise sources and illustrate the power of our method to locate the optimal
path to extinction.  Even though our examples are taken from
infectious disease models, the approach is applicable to any extinction
process or escape process.  Section~\ref{sec:opt_path} discusses stochastic
modeling in the limit of large population size, while Sec.~\ref{sec:FTLE}
describes the theory that underlies the Lyapunov exponent computations.  In
Sec.~\ref{sec:apps}, our
method is used to find the optimal path to extinction for three examples.  In
Sec.~\ref{sec:proof}, we demonstrate that the optimal path to extinction also
possesses a local maximum of the FTLE field, and conclusions are contained in Sec.~\ref{sec:conc}.

\section{Stochastic modeling in the large population limit}

\label{sec:opt_path}

To introduce the idea of constructing an optimal path in stochastic dynamical
systems, we consider the problem of extinction taken
from mathematical epidemiology.  
The stochastic formulation of the problem accounts for the
random state transitions which govern the dynamics of the system.  To
quantitatively account for the
randomness in the system, we will formulate a master equation which describes the time evolution of the probability of
finding the system in a particular state at a certain time~\citep{gar03,vanKampen_book}.

The state variables  $\bm{X}\in \mathbb{R}^{n}$ of the system describe the components
of a population, while the random state transitions which govern the dynamics are described by the transition rates $W(\bm{X},\bm{r})$, where $\bm{r}\in \mathbb{R}^{n}$ is an increment in the change of
$\bm{X}$.  Consideration of the net change in increments of the state of
the system results in the following master equation for the probability
density $\rho(\bm{X},t)$ of finding the system in state $\bm{X}$ at time $t$:
\begin{equation}
\frac{\partial\rho(\bm{X},t)}{\partial
  t}=\sum_{\bm{r}}\left[W(\bm{X}-\bm{r};\bm{r})\rho(\bm{X}-\bm{r},t)-W(\bm{X};\bm{r})\rho(\bm{X},t)\right].\label{e:MasterEquation}
\end{equation}

If the total population size of the system is $N$, the
components of the state variable can be rescaled to be fractions of the population
by letting $\bm{x}=\bm{X}/N$.  Then the general solution~\citep{Kubo1973} for
the transition probability from ${\bm x}_0$ to ${\bm x}$ in the time interval from $t_0$
to $t$
can be written as the following path integral:
\begin{equation}
P(\bm{x},t|\bm{x}_{0},t_{0})=\int d\mathcal{D}(\bm{x,p})\,{\rm exp}\left
  \{-N\int\limits_{t_{0}}^{t}ds\, \left
    [H(\bm{x}(s),\bm{p}(s),s)-\bm{p}(s)\dot{\bm{x}}(s)\right ]\right \},\label{e:probdensity}
\end{equation}
where $d\mathcal{D}(\bm{x,p})$ denotes integration over all paths.

The integral in the exponent of Eq.~(\ref{e:probdensity}) is the action,
and the Hamiltonian $H({\bm x},{\bm p};t)$ is given in general as
\begin{equation}
H({\bm x},{\bm p};t)=\sum_{\bm r} w({\bm x};{\bm r})({\rm exp}\,(\bm{p} \cdot{\bm r})-1),\label{e:Hamiltonian}
\end{equation}
where $w({\bm x};{\bm r})$ is defined as the transition rate $W$ per
individual.  The 
Hamiltonian $H$ depends both on the state of
the system ${\bm x}$ as well as the momentum ${\bm p}$, which provides an
effective force due to  stochastic fluctuations on the system.  It should be noted
that instead of using the Hamiltonian representation, one could use
the Lagrangian representation, which results in the following alternative solution:
\begin{equation}    
P(\bm{x},t|\bm{x}_{0},t_{0})=\int d\mathcal{D}(\bm{x})\,{\rm exp}\left
  \{N\int\limits_{t_{0}}^{t}ds\, L(\bm{x}(s),\dot{\bm{x}}(s),s)\right
\},\label{e:Lagrangian representation}
\end{equation}
with $L(\bm{x},\dot{\bm{x}};t)=-H(\bm{x},\bm{p};t)+\dot{{\bm x}}\cdot{\bm p}$.

The action reveals much information about the probability evolution of the system, from scaling
near bifurcation points in non-Gaussian processes to rates of extinction as a
function of epidemiological parameters~\citep{dyk90,dyscla08}.  In order to maximize the
probability of extinction, one must minimize the action.
The minimizing formulation entails finding the solution to the
Hamilton-Jacobi equation, which means that
one must solve the $2n$-dimensional system of Hamilton's equations for $\bm{x}$ and
$\bm{p}$.  The appropriate
boundary conditions of the system are such that a solution starts at a non-zero state, such as an endemic
state, and asymptotically
approaches one or more zero components of the state vector.  Therefore, a
trajectory that
is a solution to the two-point boundary value problem determines a path, which in
turn yields the probability of going from the initial state to the final state. The optimal
path to extinction is the path that minimizes the action
in either the Hamiltonian or Lagrangian representation.

To illustrate an application of the theory for a finite population, we
consider in detail an explicit example, namely the well-known problem of
extinction in a Susceptible-Infectious-Susceptible (SIS) epidemic model.  In this example, the
population consists of $S$ susceptible individuals and $I$ infectious individuals.
The population is driven via the birth rate $\mu$, which
is also equal to the death rate. The transition rates for $\bm{X}=(S,I)^T$
are given as follows:

\begin{eqnarray}
 &  & W\bigl({\bf X};(1,0)\bigr)=N{}\mu,\qquad \qquad W\bigl({\bf X};(-1,0)\bigr)=\mu X_{1},\label{eq:}\nonumber\\
 &  & W\bigl({\bf X};(0,-1)\bigr)=\mu X_{2},\qquad\quad W\bigl({\bf X};(1,-1)\bigr)=\gamma X_{2},\label{eq:SIS transition rates}\nonumber\\
 &  & W\bigl({\bf X};(-1,1)\bigr)=\beta X_{1}X_{2}/N{},
\end{eqnarray}
where $\beta$ is the mass action contact rate, $\gamma$ is the
recovery rate, and $N$ is now a parameter for the average size of the
population. For large $S,I\propto N$, fluctuations of $S$ and $I$ are small
on average. If these fluctuations are disregarded, one arrives at
the following deterministic mean-field equations for $S$ and $I$:
\begin{subequations}
\begin{flalign}
\dot{X}_{1}=&N\mu -\mu X_{1}+\gamma X_{2}-\beta X_{1}X_{2}/N{},\label{e:MeanFielda}\\
\dot{X}_{2}=&-(\mu+\gamma)X_{2}+\beta X_{1}X_{2}/N{}.\label{e:MeanFieldb}
\end{flalign}
\end{subequations}
Equations~(\ref{e:MeanFielda})-(\ref{e:MeanFieldb}) are the standard equations of the SIS model in the absence of fluctuations.
For the parameter $R_{0}=\beta/(\mu+\gamma)>1$, they have a stable, attracting 
solution $\bm{X}_{A}=N\bm{x}_{A}$ with $x_{1A}=R_{0}^{-1}$, and
$x_{2A}=1-R_{0}^{-1}$, which corresponds to endemic disease. In addition, Eqs.~(\ref{e:MeanFielda})-(\ref{e:MeanFieldb})
have an unstable stationary state (saddle point) given by $\bm{X}_{S}=N\bm{x}_{S}$
with $x_{1S}=1$ and $x_{2S}=0$, which corresponds to
the extinct, or disease-free, state.  

For $N\gg1$, the steady state distribution $\rho({\bm X})$ has a peak
at the stable state $\bm{X}_{A}$ with width $\propto N^{1/2}$. This
peak is formed over a typical relaxation time given in~\citet{dyscla08} and~\citet{sbdl09}.
However, in the process of extinction, we are interested in the probability of having a small number
of infectious individuals, which is determined by the tail of the
distribution. The distribution tail can be obtained by seeking the solution of
Eqs.~(\ref{e:MeanFielda})-(\ref{e:MeanFieldb}) in the eikonal
form~\citep{elgkam04,Doering2005,Kubo1973,wentzell76,gang87,Dykman1994d,Tretiakov2003}
given by
\begin{eqnarray}
 &  & \rho({\bm X})=\exp\,[-N\mathcal{S}({\bm x})],\qquad\qquad{\bm x}={\bm X}/N{},\nonumber \\
 &  & \rho({\bm X}+{\bm r})\approx\rho({\bm X})\exp(-{\bm p}\cdot{\bm r}),\qquad {\bm
   p}=\partial\mathcal{S}({\bm x})/\partial{\bm x},
\end{eqnarray}
where $\mathcal{S}(\bm{x)}$ is the action.

To leading order in $N{}^{-1}$, the equation for $\mathcal{S}(\bm{x})$ has
a form of the Hamilton-Jacobi equation $\dot{\mathcal{S}}=-H({\bm x},\partial_{{\bm x}}\mathcal{S};t)$,
where $\mathcal{S}$ is the effective action, and the effective Hamiltonian
is given by Eq.~(\ref{e:Hamiltonian}), with $w({\bm x};{\bm r})=N^{-1}W({\bm X};{\bm r})$
being the transition rates per individual.  The action $\mathcal{S}(\bm{x})$ can
be found from classical trajectories of the auxiliary system with
Hamiltonian $H$ that satisfy the following equations:
\begin{eqnarray}
\dot{\bm x}=\partial_{\bm p}H({\bm x},{\bm p}),\qquad
\dot{\bm p}=-\partial_{\bm x}H({\bm x},{\bm p}).\label{e:Hamilt_equations}
\end{eqnarray}

Since the maximum of the probability of extinction is found by minimizing the
action, we compute the trajectory satisfying the Hamiltonian system that has as its asymptotic
limits in time the endemic state as $t\to-\infty$ and the extinct
state as $t\to+\infty$.  The action then has the form, from
Eq.~(\ref{e:probdensity})~\citep{wentzell76,gang87,Dykman1994d,Tretiakov2003}:
\begin{equation}
\mathcal{S}(\bm{x}_{S})=\int\limits_{-\infty}^{\infty}{\bm p}\cdot\dot{\bm{x}}\,
dt,\qquad H({\bm x},{\bm p})=0.\label{e:action_stationary}
\end{equation}
In Eq.~(\ref{e:action_stationary}), the integral is calculated for a
Hamiltonian trajectory $\bigl(\bm{x}(t),{\bm p}(t)\bigr)^T$
that starts as $t\to-\infty$ at ${\bm x}\to {\bm x}_{A}, {\bm p}\to{\bm 0}$,
and arrives as $t\to\infty$ at the state ${\bm x}_{S}$. This trajectory
describes the most probable sequence of elementary events ${\bm x}\to {\bm
  x}+{\bm r}$ that
brings the system to ${\bm x}_{S}$. 

Several authors have considered how extinction rates scale with respect
to a parameter near bifurcation
points~\citep{Doering2005,kammee08,Kamenev2008b,dyscla08} when the distance to
the bifurcation point is small and the dynamics is very slow.  For an epidemic model,
this means that the reproductive rate of infection is greater than
but very close to one.  However, most real diseases have reproductive rates of
infection greater than $1.5$~\citep{Anderson91}, which translates into faster growth rates from
the extinct state.
In general, in order to get analytic scaling results, one must compute
the optimal path using either the Hamiltonian or Lagrangian equations
of motion.  However, far from bifurcation points, one is seldom able to
perform the required analysis or computation.

Additionally, the computation of the optimal path involves the use of a
numerical shooting method~\citep{kammee08}.  The Hamiltonian or Lagrangian
representation of an $n$-dimensional dynamical system lies in $2n$-dimensional
space.  Therefore, for even relatively low-dimensional dynamical systems, the
use of a shooting method to find the optimal path to extinction can be quite
problematic.  In the next section, we demonstrate how to evolve naturally to
the optimal path using a dynamical systems approach.

\section{Finite-Time Lyapunov Exponents}\label{sec:FTLE}
We consider a velocity field ${\bm{v}}:\mathbb{R}^{2n}\times
I\rightarrow\mathbb{R}^{2n}$ given by the Hamiltonian field in Eq.~(\ref{e:Hamilt_equations})
which is defined over the time interval $I=[t_{i},t_{f}]\subset\mathbb{R}$
and the following system of equations: \begin{subequations} \begin{flalign}
 & \dot{{\bm{y}}}(t;t_{i},{\bm{y}}_{0})={\bm{v}}({\bm{y}}(t;t_{i},{\bm{y}}_{0}),t),\label{e:xdot}\\
 & {\bm{y}}(t_{i};t_{i},{\bm{y}}_{0})={\bm{y}}_{0},\label{e:xIC}\end{flalign}
 \end{subequations} where ${\bm{y}}=(\bm{x},{\bm p})^T\in\mathbb{R}^{2n}$, ${\bm{y}}_{0}\in\mathbb{R}^{2n}$,
and $t\in I$.

Such a continuous time dynamical system has quantities, known as Lyapunov
exponents, which are associated with the trajectory of the system
in an infinite time limit. The Lyapunov exponents measure the growth
rates of the linearized dynamics about the trajectory. To find the
finite-time Lyapunov exponents (FTLE), one computes the Lyapunov exponents
on a restricted finite time interval.   For each initial condition, the
exponents provide a measure of its sensitivity to small perturbations.
Therefore, the FTLE is
a measure of the local sensitivity to initial data.  For the purpose of completeness,
we briefly recapitulate the derivation of the FTLE. Details regarding
the derivation along with the appropriate smoothness assumptions can
be found in~\citet{hall00,hall01,hall02},~\citet{shlema05},~\citet{leshma07}, and~\citet{brawig09}.

The integration of Eqs.~(\ref{e:xdot})-(\ref{e:xIC}) from the initial
time $t_{i}$ to the final time $t_{i}+T$ yields the flow map $\phi_{t_{i}}^{t_{i}+T}$
which is defined as follows: \begin{equation}
\phi_{t_{i}}^{t_{i}+T}:{\bm{y}}_{0}\mapsto\phi_{t_{i}}^{t_{i}+T}({\bm{y}}_{0})={\bm{y}}(t_{i}+T;t_{i},{\bm{y}}_{0}).\label{e:map}\end{equation}
 Then the FTLE can be defined as \begin{equation}
\sigma({\bm{y}},t_{i},T)=\frac{1}{|T|}\ln{\sqrt{\lambda_{{\rm max}}(\Delta)}},\label{e:sigma}\end{equation}
 where $\lambda_{{\rm max}}(\Delta)$ is the maximum eigenvalue of
the right Cauchy-Green deformation tensor $\Delta$, which is given
as follows: \begin{equation}
\Delta({\bm{y}},t_{i},T)=\left(\frac{d\phi_{t_{i}}^{t_{i}+T}({\bm{y}}(t))}{d{\bm{y}}(t)}\right)^{*}\left(\frac{d\phi_{t_{i}}^{t_{i}+T}({\bm{y}}(t))}{d{\bm{y}}(t)}\right),\label{e:delta}\end{equation}
 with {*} denoting the adjoint.

For a given ${\bm{y}}\in\mathbb{R}^{2n}$ at an initial time $t_{i}$,
Eq.~(\ref{e:sigma}) gives the maximum finite-time Lyapunov exponent
for some finite integration time $T$ (forward or backward), and provides
a measure of the sensitivity of a trajectory to small perturbations.

\begin{figure}[t!]
\begin{center}
\includegraphics[width=100mm]{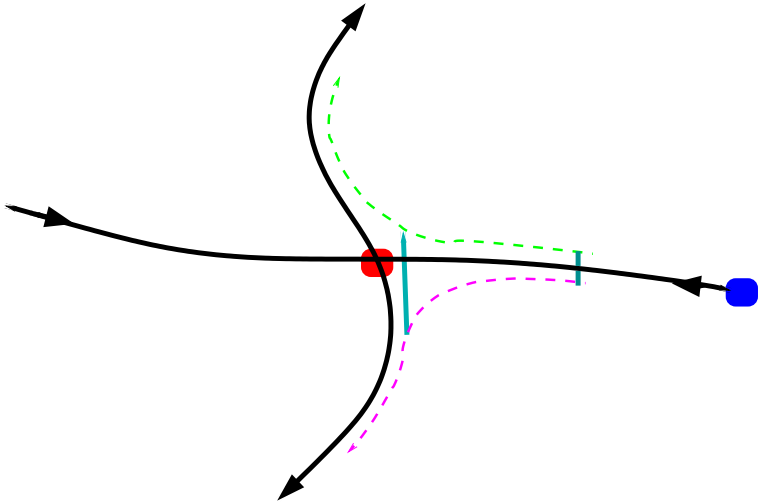} %
\caption{\label{fig:hyperbolic}(Color online) Schematic showing the path from
  the endemic state (blue) to the extinct state (red).  The optimal path leaves the
  endemic point along an unstable manifold and connects to the extinct state
  along a stable manifold.  The magenta and green dashed lines represent
  trajectories on either side of the optimal path. The initial starting distance
  between trajectories near the endemic state expands exponentially in forward
  time (shown by the cyan lines).  Locally, this demonstrates that the finite-time Lyapunov measure of
  sensitivity with respect to initial data is maximal along the optimal
  path.  This is evident in the ridges observed in the evolution of the
  exponents.}
\end{center}
\end{figure}

The FTLE field given by $\sigma({\bm{y}},t_{i},T)$ can be shown to exhibit ``ridges'' of local maxima in phase
space.  The ridges
of the field indicate the location of attracting (backward time FTLE
field) and repelling (forward time FTLE field) structures.  In two-dimensional (2D) space, the ridge is a curve which locally maximizes
the FTLE field so that transverse to the ridge one finds the FTLE to
be a local maximum.  What is remarkable is that the FTLE ridges correspond to
the optimal path trajectories, which is shown heuristically in Sec.~\ref{sec:proof}.  The basic idea is that since the optimal path is inherently
unstable and observed only through many realizations of stochastic
experiments, the FTLE shows that locally, the path is also the most sensitive
to initial data.  Figure~\ref{fig:hyperbolic} shows a schematic that
demonstrates why the optimal path has a local maximum to sensitivity.  If one chooses
an initial point on either side of the path near the endemic state, the two trajectories
will separate exponentially in time.  This is due to the fact that both extinct
and endemic states are unstable, and the connecting trajectory defining the
path is unstable as well.  Any initial points starting near the optimal path
will leave the neighborhood in short time.

\section{Finding the Optimal Path to Extinction Using FTLE}

\label{sec:apps}
We now apply our theory of dynamical sensitivity to the problem of locating
optimal paths to extinction for several examples.  We consider the case of
internal fluctuations, where the noise is not known a priori, as well as the
case of external noise, where the noise is specified.  In each case, the interaction of
the noise and state of the systems begin with a description of the Hamiltonian
(or Lagrangian) to describe the unstable flow.  Then the corresponding
equations of motion are used to compute the ridges corresponding to maximum FTLE, which
in turn correspond to the optimal extinction paths~\citep{sfbs2010}.

\subsection{Example 1 - Extinction in a Branching-Annihilation Process}

\label{sec:birth_death} For an example of a system with intrinsic noise
fluctuations which has an analytical solution, we consider extinction in the stochastic branching-annihilation
process \begin{equation}
A\xrightarrow{\lambda}2A\quad{\rm and}\quad2A\xrightarrow{\mu}\emptyset,\label{e:bd}\end{equation}
 where $\lambda$ and $\mu>0$ are constant reaction rates~\citep{elgkam04,askame08}.
Equation~(\ref{e:bd}) is a single species birth-death process and
can be thought of as a simplified form of the Verhulst logistic model
for population growth~\citep{nase01}.

The stochastic process given by Eq.~(\ref{e:bd}) contains intrinsic
noise which arises from the randomness of the reactions and the fact
that the population consists of discrete individuals. This intrinsic
noise, which can generate a rare sequence of events that cause the
system to evolve to the empty state, can be accounted for using a
master equation approach. The probability $P_{n}(t)$ to observe,
at time $t$, $n$ individuals is governed by the following master
equation: \begin{equation}
\dot{P}_{n}=\frac{\mu}{2}\left[(n+2)(n+1)P_{n+2}-n(n-1)P_{n}\right]+\lambda\left[(n-1)P_{n-1}-nP_{n}\right].\label{e:bd_master}\end{equation}

Using the formalism of~\citet{askame08}, Eq.~(\ref{e:bd_master})
is recast as the following exact evolution equation for $G(\rho,t)$:
\begin{equation}
\frac{\partial G}{\partial t}=\frac{\mu}{2}(1-\rho^{2})\frac{\partial^{2}G}{\partial\rho^{2}}+\lambda\rho(\rho-1)\frac{\partial G}{\partial\rho},\label{e:bd_master_2}\end{equation}
 where $G$ is a probability generating function given by \begin{equation}
G(\rho,t)=\sum\limits _{n=0}^{\infty}\rho^{n}P_{n}(t),\label{e:gen_fn}\end{equation}
 and where $\rho$ is an auxiliary variable.

We substitute the eikonal ansatz $G(\rho,t)={\rm exp}[-\mathcal{S}(\rho,t)]$,
where $\mathcal{S}$ is the action, into Eq.~(\ref{e:bd_master_2})
and neglect the higher-order $\partial^{2}\mathcal{S}/\partial\rho^{2}$
term. This results in a Hamilton-Jacobi equation for $\mathcal{S}(\rho,t)$.
By introducing a conjugate coordinate $q=-\partial\mathcal{S}/\partial\rho$
and by shifting the momentum $p=\rho-1$, then one arrives at the
following Hamiltonian: \begin{equation}
H(q,p)=\left(\lambda(1+p)-\frac{\mu}{2}(2+p)q\right)qp.\label{e:bd_Ham}\end{equation}
 Hamilton's equations are therefore given as: \begin{subequations}
\begin{flalign}
\dot{q}= & \frac{\partial H}{\partial p}=q[\lambda(1+2p)-\mu(1+p)q],\label{e:bd_qdot}\\
\dot{p}= & -\frac{\partial H}{\partial q}=p[\mu(2+p)q-\lambda(1+p)].\label{e:bd_pdot}\end{flalign}
 \end{subequations}

The Hamiltonian given by Eq.~(\ref{e:bd_Ham}) has three zero-energy
curves. The first is the mean-field zero-energy line $p=0$, which
contains two hyperbolic points given as $h_{1}=(q,p)=(\lambda/\mu,0)$
and $h_{0}=(q,p)=(0,0)$. The second is the extinction line $q=0$,
which contains another hyperbolic point given as $h_{2}=(q,p)=(0,-1)$.
The third zero-energy curve is non-trivial and is given as follows:
\begin{equation}
q=\frac{2\lambda(1+p)}{\mu(2+p)}.\label{e:bd_op}\end{equation}
The segment of the curve given by Eq.~(\ref{e:bd_op}) which lies
between $-1\leq p\leq0$ corresponds to a heteroclinic trajectory.
 As $t$ approaches $-\infty$,  the trajectory exits the hyperbolic point $h_{1}$ along
its unstable manifold and enters the hyperbolic point
$h_{2}$ along its stable manifold  as  $t$ approaches $\infty$. 
This heteroclinic trajectory is
the optimal path to extinction and describes the most probable (rare)
sequence of events which evolves the system from a quasi-stationary
state to extinction~\citep{askame08}.

\begin{figure}[t!]
\begin{center}
 \includegraphics[scale=0.71]{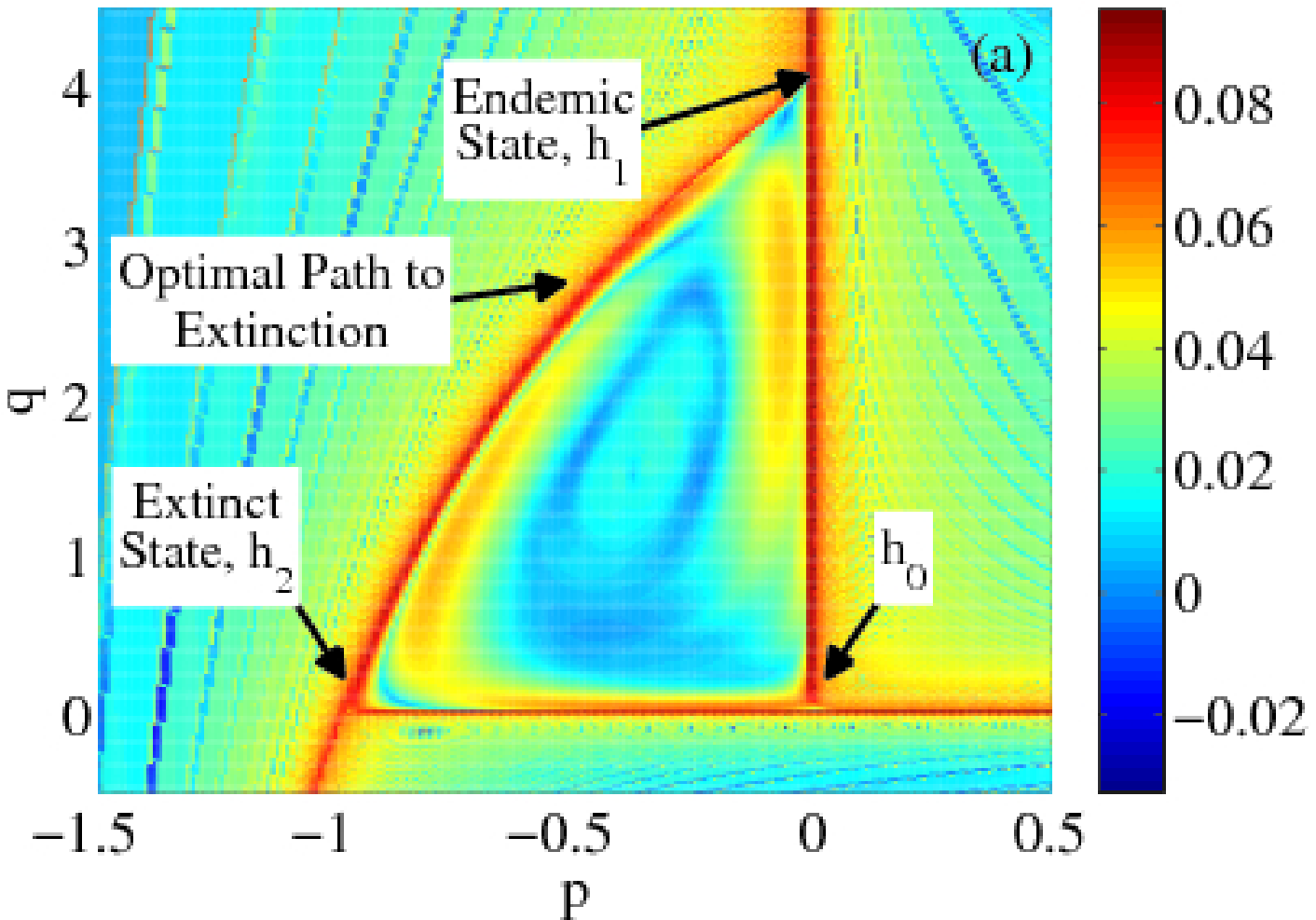}\\
 \includegraphics[scale=0.71]{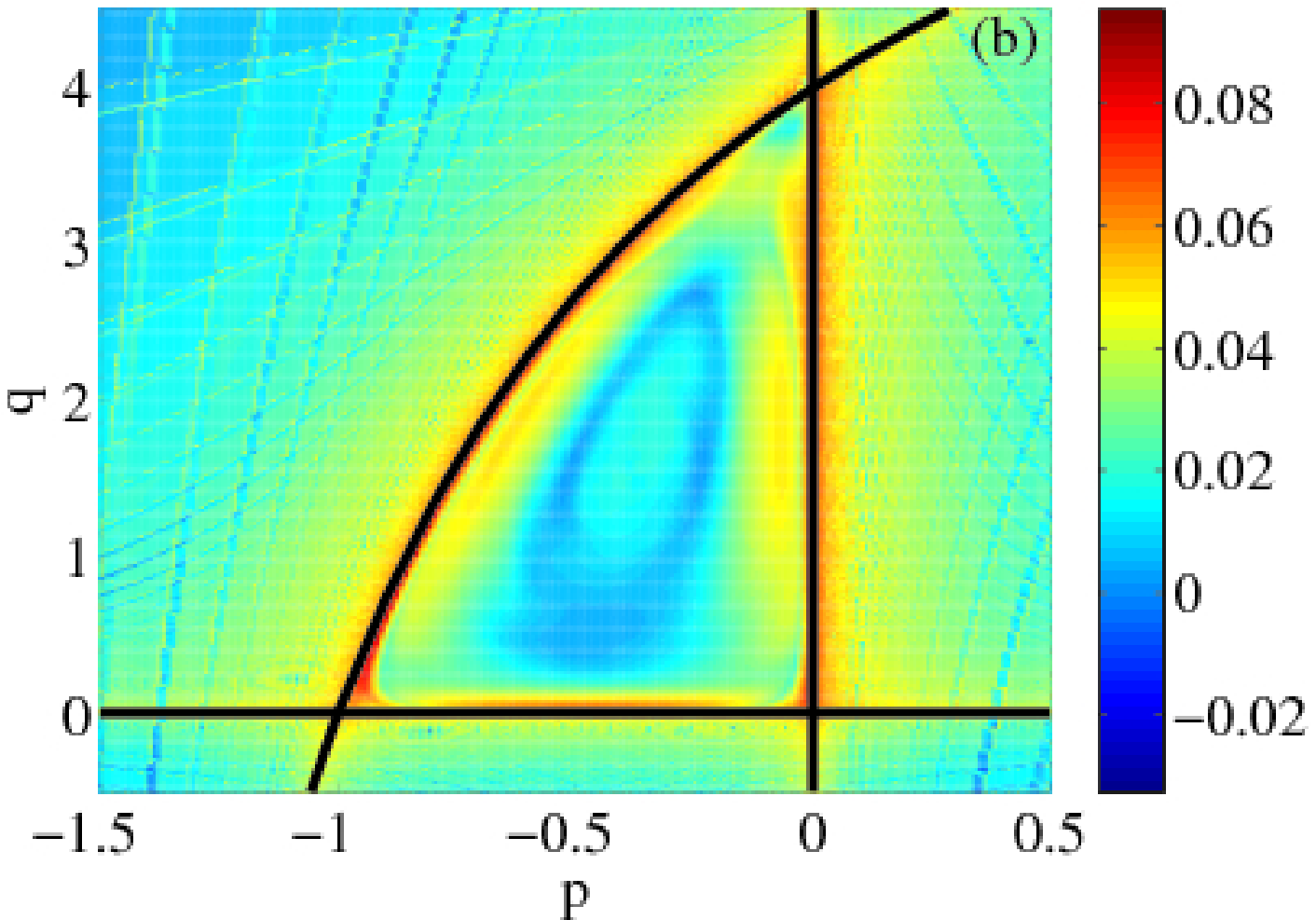}
\caption{\label{fig:FTLE_bd}(Color online) FTLE flow fields computed using
Eqs.~(\ref{e:bd_qdot})-(\ref{e:bd_pdot}) with $\lambda=2.0$ and
$\mu=0.5$. The integration time is $T=6$ with an integration step
size of $t=0.1$ and a grid resolution of $0.01$ in both $q$ and
$p$. (a) Forward FTLE field with the optimal path to extinction given
by the LCS. (b) Average of the forward and backward FTLE fields with the three zero-energy
curves given by the LCS and overlaid with the analytical solution
of these curves given by $p=0$, $q=0$, and Eq.~(\ref{e:bd_op}).  Note that the
averaging affects only the value of the FTLE and not the structure of the FTLE field.}
\end{center}
\end{figure}

To show that the FTLE evolves to the optimal path, we calculate the FTLE field using
Eqs.~(\ref{e:bd_qdot})-(\ref{e:bd_pdot}). Figure~\ref{fig:FTLE_bd}(a)
shows the forward FTLE plot computed using Eqs.~(\ref{e:bd_qdot})-(\ref{e:bd_pdot})
for $T=6$, with $\lambda=2.0$ and $\mu=0.5$. In Fig.~\ref{fig:FTLE_bd}(a),
one can see that the optimal path to extinction is given by the ridge
associated with the maximum FTLE.

Not all of the attracting structures are shown in
Fig.~\ref{fig:FTLE_bd}(a) because the maximum forward time FTLE identifies
repelling structures where nearby initial conditions diverge.  The other attracting structures may be found by computing the backward FTLE field.
By overlaying 
the forward and backward FTLE fields, one can see in
their entirety all three zero-energy curves including the optimal
path to extinction in Fig.~\ref{fig:FTLE_bd}(b). Also shown in Fig.~\ref{fig:FTLE_bd}(b)
are the analytical solutions to the three zero-energy curves given
by $p=0$, $q=0$, and Eq.~(\ref{e:bd_op}). There is excellent agreement
between the analytical solutions of all three curves and the LCS which
are found through numerical computation of the FTLE flow fields.

\subsection{Example 2 - SIS Epidemic Model - External Fluctuations}\label{sec:SIS_L}
As another general application of the extinction theory for finite
populations, we consider the well-known problem of extinction in a Susceptible-Infectious-Susceptible
(SIS) epidemiological model.  The SIS model is given by the following system of equations:
\begin{subequations} \begin{flalign}
\dot{S}= & \mu-\mu S+\gamma I-\beta IS,\label{e:S}\\
\dot{I}= & -(\mu+\gamma)I+\beta IS,\label{e:I}\end{flalign}
 \end{subequations} where $\mu$ represents a constant birth and
death rate, $\beta$ represents the contact rate, and $\gamma$ denotes
the rate of recovery.  If we assume that the total population size
is constant and can be normalized to $S+I=1$, then Eqs.~(\ref{e:S})-(\ref{e:I})
can be rewritten as the following one-dimensional (1D) equation: \begin{equation}
\dot{I}=-(\mu+\gamma)I+\beta I(1-I).\label{e:1DSIS}\end{equation}

\begin{figure}[t!]
\begin{center}
\includegraphics[scale=0.72]{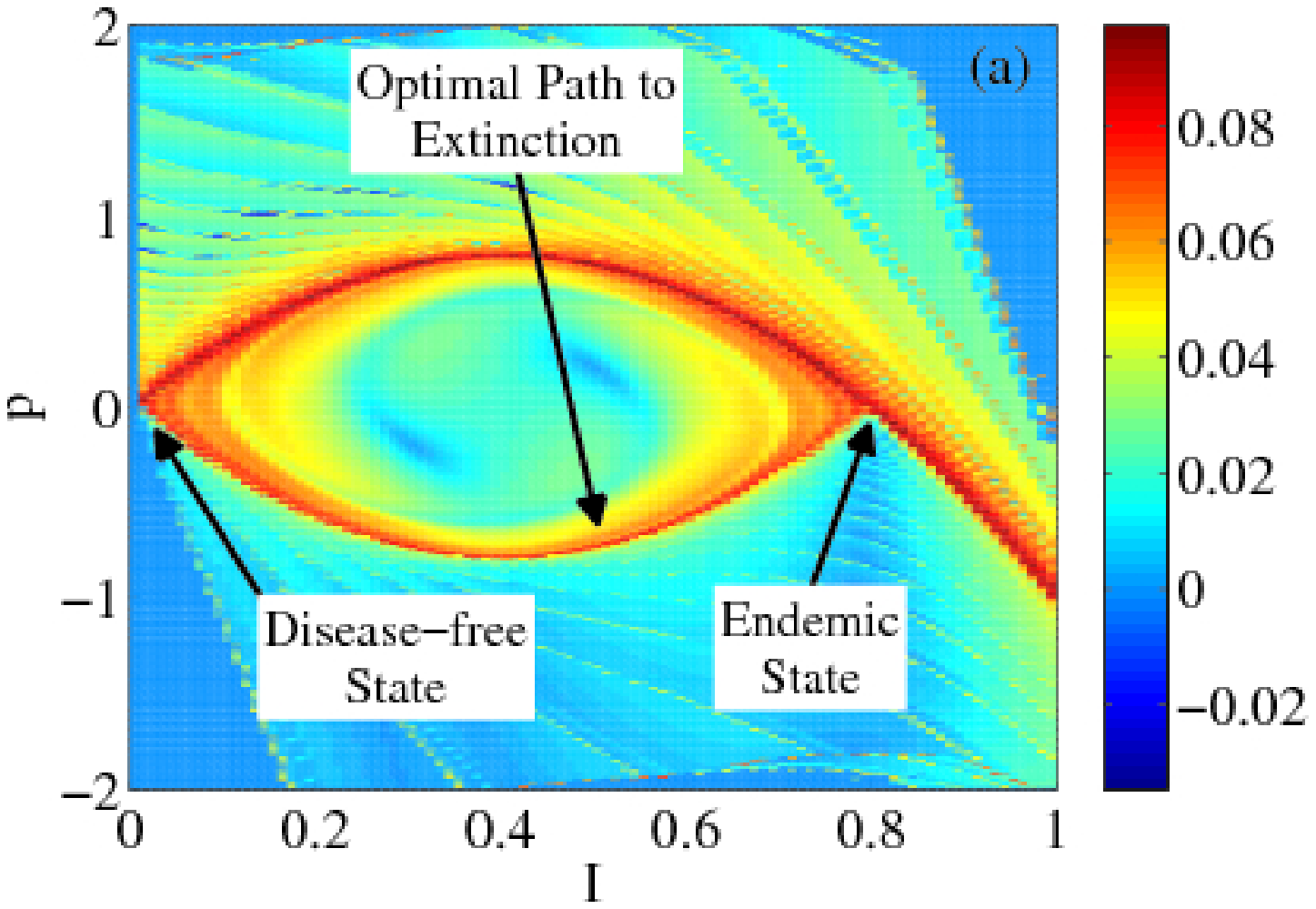} \\
\includegraphics[scale=0.72]{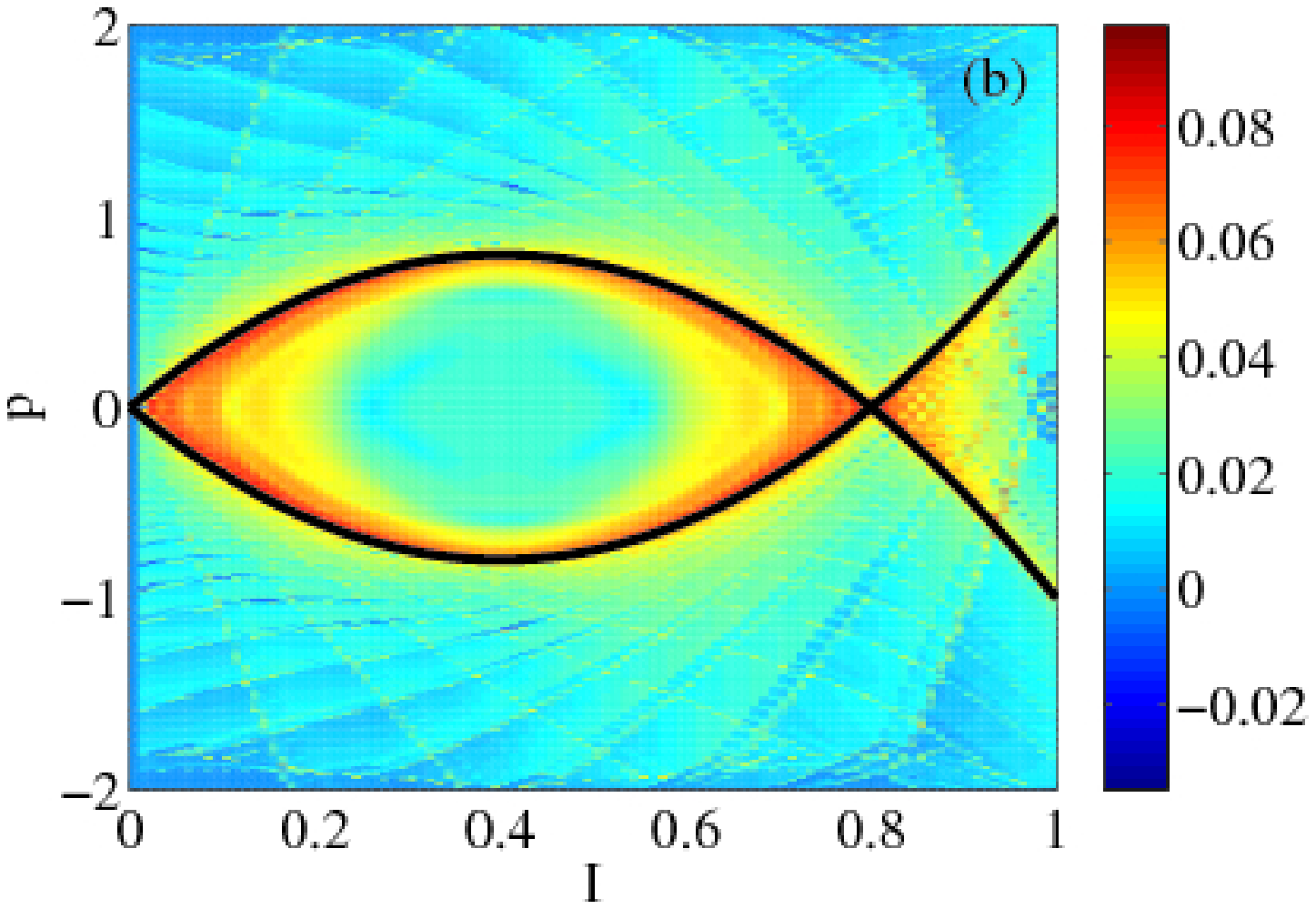}
\caption{\label{fig:FTLE_Lag}(Color online) FTLE flow fields computed using
Eqs.~(\ref{e:Idot})-(\ref{e:pdot}) with $\beta=5.0$ and $\kappa=1.0$.
The integration time is $T=5$ with an integration step size of $t=0.1$
and a grid resolution of $0.01$ in both $I$ and $p$. (a) Forward
FTLE field with the optimal path to extinction given by the LCS. (b)
Average of the forward and backward FTLE fields with the optimal path to extinction
and its counterpart optimal path given by the LCS and overlaid with
the analytical solution of the optimal paths given by Eq.~(\ref{e:opt_path}).
Note that the
averaging affects only the value of the FTLE and not the structure of the FTLE field.}
\end{center}
\end{figure}

The stochastic version of Eq.~(\ref{e:1DSIS}) is given as \begin{equation}
\dot{I}=-(\mu+\gamma)I+\beta I(1-I)+\eta(t)=F(I)+\eta(t),\label{e:1DSIS_stoch}\end{equation}
 where $\eta(t)$ is uncorrelated Gaussian noise with zero mean and models
 random migration to and from another
 population~\citep{Alonso2006,Doering2005}.  Equation~(\ref{e:1DSIS_stoch})
has two equilibrium points given by $I=0$ (corresponding to the disease-free
state) and $I=1-(\mu+\gamma)/\beta$ (corresponding to the endemic
state). One can use the Euler-Lagrange equation of motion to find
the optimal path of extinction from the endemic state to the disease-free
state, where the Lagrangian is determined by Eq.~(\ref{e:1DSIS_stoch})
and is given as follows: \begin{equation}
L(I,\dot{I})=[\eta(t)]^{2}=[\dot{I}-F(I)]^{2}.\end{equation}
 Computation of the Euler-Lagrange equation gives the following: \begin{equation}
F(I)F^{\prime}(I)-\ddot{I}=0.\label{e:EL}\end{equation}
 If one multiplies Eq.~(\ref{e:EL}) by $\dot{I}$ followed by an
integration with respect to $t$, then one obtains \begin{equation}
\frac{F(I)^{2}}{2}-\frac{\dot{I}^{2}}{2}=E,\end{equation}
 where $E$ is an arbitrary constant. Using the fact that the optimal
path passes through the two equilibrium points stated above, then
one finds that the optimal path to extinction (as well as its counterpart
path from the disease-free state to the endemic state) is given by
the following equation: \begin{equation}
\dot{I}=\pm F(I).\label{e:opt_path}\end{equation}

As in the first example, one can numerically compute the optimal path to extinction
using the FTLE. In this example we calculate the FTLE field using
the following 2D system which is equivalent to Eq.~(\ref{e:EL}):
\begin{subequations} \begin{flalign}
\dot{I}= & p,\label{e:Idot}\\
\dot{p}= & F(I)F^{\prime}(I)=(\beta I(1-I)-\kappa I)(\beta(1-2I)-\kappa),\label{e:pdot}\end{flalign}
 \end{subequations} where $\kappa=\mu+\gamma$. Figure~\ref{fig:FTLE_Lag}(a)
shows the forward FTLE plot computed using Eqs.~(\ref{e:Idot})-(\ref{e:pdot})
for $T=5$, with $\beta=5.0$ and $\kappa=1.0$. In Fig.~\ref{fig:FTLE_Lag}(a),
one can see that the optimal path from the endemic state to the disease-free
state is given by the ridge associated with the locally maximal FTLE.

One also can find the optimal path from the disease-free state to
the endemic state by computing the backward FTLE. By overlaying the
forward and backward FTLE fields, one can see the optimal path to
extinction along with its counterpart optimal path in Fig.~\ref{fig:FTLE_Lag}(b).
Also shown in Fig.~\ref{fig:FTLE_Lag}(b) are the two analytical
solutions to the optimal path to extinction and its counterpart optimal
path which are given by Eq.~(\ref{e:opt_path}). There is excellent
agreement between the analytical solutions to the two optimal paths
and the ridges which are found through numerical computation of the
FTLE flow fields.

\subsection{Example 3 - SIS Epidemic Model - Internal Fluctuations}

\label{sec:SIS_H} We now consider the 1D stochastic (internal) version of
the SIS epidemic model given by Eq.~(\ref{e:1DSIS}). The probability
$P_{n}(t)$ to observe, at time $t$, $n$ infectious individuals
is governed by the following master equation: \begin{equation}
\dot{P}_{n}=(\mu+\gamma)[(n+1)P_{n+1}-nP_{n}]+\beta[(n-1)(1-(n-1))P_{n-1}-n(n-1)P_{n}].\label{e:master_eq}\end{equation}
 Using the formalism of~\citet{gang87}, one then has the following
Hamiltonian associated with Eq.~(\ref{e:master_eq}): \begin{equation}
H(I,p)=(\mu+\gamma)I(e^{-p}-1)+\beta I(1-I)(e^{p}-1),\end{equation}
 and Hamilton's equations are therefore given as: \begin{subequations}
\begin{flalign}
\dot{I}= & \frac{\partial H}{\partial p}=-(\mu+\gamma)Ie^{-p}+\beta I(1-I)e^{p},\label{e:Idot_HE}\\
\dot{p}= & -\frac{\partial H}{\partial I}=-(\mu+\gamma)(e^{-p}-1)+\beta(e^{p}-1)(2I-1).\label{e:pdot_HE}\end{flalign}
 \end{subequations}

\begin{figure}[h!]
\begin{center}
 \includegraphics[scale=0.74]{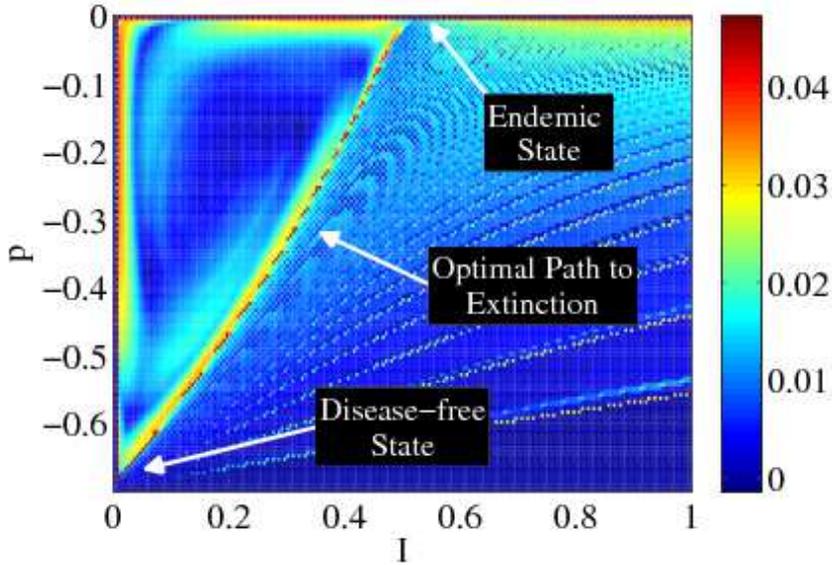}

\caption{\label{fig:FTLE_Ham}(Color online) FTLE flow field computed using
Eqs.~(\ref{e:Idot_HE})-(\ref{e:pdot_HE}) with $\beta=2.0$ and
$\kappa=1.0$. The integration time is $T=10$ with an integration
step size of $t=0.1$ and a grid resolution of $0.005$ in both $I$
and $p$. The optimal path to extinction is given by the LCS.}
\end{center}
\end{figure}

Although there is no analytical solution for the optimal path to extinction
for Eqs.~(\ref{e:Idot_HE})-(\ref{e:pdot_HE}), we can once again
determine the optimal path by computing the FTLE flow field associated
with this system. Figure~\ref{fig:FTLE_Ham} shows the forward FTLE
plot computed using Eqs.~(\ref{e:Idot_HE})-(\ref{e:pdot_HE}) for
$T=10$, with $\beta=2.0$ and $\kappa=1.0$. As we have seen previously,
the optimal path to extinction from the endemic state to the disease-free
state is given by the ridge associated with the locally maximal FTLE.

\section{Maximal Sensitive Dependence to Initial Data Near the Optimal Path}\label{sec:proof}

The main heuristic argument of this section is to show that the path
which maximizes the probability of extinction also has a finite-time
Lyapunov exponent (FTLE) that attains its local maximum on the path.
We consider a general system of equations and show how the maximum
unstable direction along the optimal path to extinction governs the
local hyperbolic dynamics.

From the Hamiltonian or Lagrangian equations of motion, the process
which leads to extinction consists of a trajectory which emanates
from a stable steady state ${\bm{x}}_{a}\in{\mathbb{R}}^{n}$ and
approaches the extinct state ${\bm{x}}_{s}\in{\mathbb{R}}^{n}$. Since
the stable and extinct states are both regular saddles (or unstable
foci) in the variational formulations, they both have hyperbolic structure.
Moreover, every point along the trajectory connecting the two states
as $t\rightarrow\pm\infty$ is assumed to possess a local hyperbolic
structure. As an example, consider the Langevin problem having a vector
field of position ${\bm{V}}:{\mathbb{R}}^{n}\rightarrow{\mathbb{R}}^{n}$,
which has the associated Lagrangian $L({\bm{x}},\dot{\bm{x}})=\left\Vert \dot{{\bm{x}}}-{\bm{V}}({\bm{x}})\right\Vert {}^{2}/2$
to describe the action. Converting to a Hamiltonian formulation leads
one to the following $2n$-dimensional equations of motion and Hamiltonian:
\begin{subequations} 
\begin{flalign}
 & \dot{\bm{x}}={\bm{p}}+{\bm{V}}({\bm{x}}),\label{eq:H equations of motion 1d_a}\\
 & \dot{\bm{p}}=-{\bm{V}}^{\prime}({\bm{x}}){\bm{p}},\label{eq:H equations of motion 1d_b}\\
 & H({\bm{x}},{\bm{p}})=\frac{\left\Vert {\bm{p}}\right\Vert
   ^{2}}{2}+{\bm{p}}\cdot{\bm{V}}({\bm{x}}),\label{eq:H equations of motion
   1d_c}
\end{flalign}
\end{subequations}
where $\bm{V}^{\prime}(\bm{x})\equiv\frac{\partial {\bm{V}}(\bm{x})}{\partial
  {\bm x}}$ is the
Jacobian matrix evaluated at ${\bm x}$. 

It is immediate from Eqs.~(\ref{eq:H equations of motion 1d_a})-(\ref{eq:H equations of motion 1d_c})
that $\left\{ ({\bm{x}},{\bm{p}})\,|\,\,{\bm{p}}={\bm{0}}\right\} $
is an invariant manifold. In addition, the optimal path must lie along
the $H({\bm{x}},{\bm{p}})=0$ surface, which means that in addition
to the ${\bm{p}}={\bm{0}}$ manifold, the zero surface includes $\left\{ ({\bm{x}},{\bm{p}})\,|\,\,{\bm{p}}=-2{\bm{V}}({\bm{x}})\right\} $.

To clarify the direction along the optimal path as well as
the local geometry, we make the following assumptions regarding ${\bm{V}({\bm{x}})}$: 
\begin{enumerate}
\item ${\bm{V}}({\bm{x}})$ is smooth, 
\item ${\bm{V}}({\bm{x}_{a}})={\bm{V}}({\bm{x}}_{s})={\bm{0}}$, 
\item ${\bm{V}}^{\prime}({\bm{x}}_{a})$ has eigenvalues with negative real
parts, and ${\bm{V}}^{\prime}({\bm{x}}_{s})$ has at least one eigenvalue
with positive real part. 
\end{enumerate}
Items 2 and 3 imply that ${\bm{x}}_{a}$ is an attracting steady state
and ${\bm{x}}_{s}$ is an unstable steady state in the deterministic
dynamical system. We now assume that the optimal path must satisfy
$\mathop{\lim}\limits _{t\to+\infty}({\bm{x}}(t),{\bm{p}}(t))=({\bm{x}}_{s},{\bm{0}})$,
while $\mathop{\lim}\limits _{t\to-\infty}({\bm{x}}(t),{\bm{p}}(t))=({\bm{x}}_{a},{\bm{0}})$.
Since $H({\bm{x}}(t),{\bm{p}}(t))=0$ along the path, the limits provide
direction along the optimal path.

The optimal path lies on the curve 
\begin{equation*}
C_{({\bm{x}},{\bm{p}})}=\left\{
  t\in(-\infty,\infty)\,|\,{\bm{p}}(t)=-2{\bm{V}}({\bm{x}}(t))\right\},
\end{equation*}
and ${\bm{p}}={\bm{0}}$ corresponds to the zero fluctuation case.
We shift the optimal path to the origin 
by using the following $2n$-dimensional transformation: \begin{subequations}
\begin{flalign}
 & {\bm{u}}={\bm{x}},\label{e:-1a}\\
 & {\bm{w}}={\bm{p}}+2{\bm{V}}({\bm{x}}),\label{e:-1b}\\
 & \hat{H}({\bm{u}},{\bm{w}})=\frac{\left\Vert {\bm{w}}\right\Vert ^{2}}{2}-{\bm{w}}\cdot{\bm{V}}(\bm{u}).\label{e:-1c}\end{flalign}
 \end{subequations} The new equations of motion are now: \begin{subequations}
\begin{flalign}
 & \dot{\bm{u}}=\partial\hat{H}/\partial{\bm{w}}={\bm{w}}-{\bm{V}}({\bm{u}}),\label{e:-2a}\\
 & \dot{\bm{w}}=-\partial\hat{H}/\partial{\bm{u}}={\bm{V}}^{\prime}({\bm{u}}){\bm{w}}.\label{e:-2b}\end{flalign}
 \end{subequations}\\
 The optimal path now is described by the curve 
\begin{equation*}
C_{({\bm{u}},{\bm{w}})}=\left\{
  t\in(-\infty,\infty)\,|\,{\bm{w}}(t)={\bm{0}},\dot{\bm{u}}(t)=-{\bm{V}}({\bm{u}}(t))\right\},
\end{equation*}
while the zero fluctuation case given by ${\bm{p}}={\bm{0}}$ now
corresponds to ${\bm{w}}=2{\bm{V}}({\bm{u}})$. 

The linearized variation along the optimal path $C_{({\bm{u}},{\bm{w}})}$
is given by the following matrix initial value problem from Eqs.~(\ref{e:-2a})-(\ref{e:-2b}):
\begin{equation}
\dot{{\bm{X}}}=\left[\begin{array}{cc}
-{\bm{V}}^{\prime}({\bm{u}}(t)) & {\bm{I}}_{n}\\
{\bm{0}} & {\bm{V}}^{\prime}({\bm{u}}(t))\end{array}\right]{\bm{X}}\equiv{\bm{J}}({\bm{u}}(t),{\bm{0}}){\bm{X}},\quad{\bm{X}}({\bm{0}})={\bm{I}}.\label{eq:LVE}\end{equation}

For a fixed time $t_{0}$ such that ${\bm{u}}(t_{0})={\bm{u}}_{0}$,
the local eigenvalues of Eq.~(\ref{eq:LVE}) are given by the eigenvalues
of $\pm{\bm{V}}^{\prime}({\bm{u}}_{0})$ and are assumed to have non-zero
real part for any $({\bm{u}}_{0},{\bm{0}})\in C_{({\bm{u}},{\bm{w}})}$.
Thus the optimal path is hyperbolic at every point.  We also
suppose the existence of a local coordinate system on the path so
that there exists a set of linearly independent directions pointwise.

The solution to the linear variational equation about $({\bm{u}},{\bm{w}})=({\bm{u}}_{0},{\bm{0}})$
for $0<t\ll1$ is given by ${\bm{X}}(t)\approxeq\exp{(t{\bm{J}}({\bm{u}}_{0},{\bm{0}}))}$.
We assume for simplicity that the eigenvalues of ${\bm{V}}'(u_{0})$
have algebraic multiplicity of one. The eigenvalues of ${\bm{J}}({\bm{u}}_{0},{\bm{0}})$
are given by $\{\pm\lambda_{i}\}_{i=1}^{n}$, where $\lambda_{i}$
are eigenvalues of ${\bm{V}}^{\prime}({\bm{u}}_{0})$. To examine
the dynamic instability that dominates locally, let $\lambda_{{\rm max}}$
denote the eigenvalue with largest real part and be such that $Re(\lambda_{max})>0$.
Since $-\lambda_{{\rm max}}$ has the most negative real part, its
eigendirection denotes the strongest contracting direction.

For any $({\bm{u}}_{0},{\bm{0}})$ on the path, the existence
of a set of linearly independent eigensolutions of ${\bm{J}}({\bm{u}}_{0},{\bm{0}})$
implies there is a transformation ${\bm{X}}={\bm{P}}{\bm{Y}}$, with
$\bm{P}\in{\mathbb{L}}^{2n\times2n},$ that diagonalizes the linear
variational equation given by Eq.~(\ref{eq:LVE}).  Without
loss of generality, we can also assume that the diagonal consists
of ordered descending eigenvalues based on the real part. Therefore, the linear variational system has the form

\begin{equation}
\dot{\bm{Y}}=\left[\begin{array}{cccccc}
\lambda_{{\rm max}}\\
 & \ddots\\
 &  & \lambda_{n}\\
 &  &  & -\lambda_{n}\\
 &  &  &  & \ddots\\
 &  &  &  &  & -\lambda_{{\rm max}}\end{array}\right]{\bm{Y}}.\label{eq:LVE2}\end{equation}

For any initial value, the solution to Eq.~(\ref{eq:LVE2}) is 
\begin{flalign}
{\bm{x}}_{p}(t;{\bm{x}}_{0}) & =(x_{1}(t),x_{2}(t),\cdots,x_{2n}(t))\\
 & =(e^{\lambda_{{\rm
       max}}t}x_{10},e^{\lambda_{2}t}x_{20},\cdots,e^{\lambda_{n}t}x_{n0},\nonumber\\
&\qquad \quad e^{-\lambda_{n}t}x_{(n+1)0},\cdots,e^{-\lambda_{2}t}x_{(2n-1)0},e^{-\lambda_{{\rm max}}t}x_{2n0}).\label{eq:solution_to_LVE2}
\end{flalign}

To show that the FTLE takes it maximum along the path, we notice that
any point along the path is hyperbolic with a saddle structure. Therefore,
we consider an arbitrary initial condition lying within a small domain
containing the origin. Since almost any initial condition hits the
boundary of the domain in finite time due to the saddle structure
of the origin, we use the escape time as the final time for the FTLE.
The definition we use of the FTLE is the direct comparison of the
distance between two close trajectories as follows: \begin{equation}
\sigma(t;{\bm{x}}_{0})=\frac{1}{t}\ln{(||{\bm{x}}_{p}(t;{\bm{x}}_{0}+{\bm{\epsilon}})-{\bm{x}}_{p}(t;{\bm{x}}_{0})||)},\label{eq:FTLE difference defn}\end{equation}
 where ${\bm{\epsilon}}\in{\mathbb{R}}^{2n}$.

Defining the domain to be the $2n$-dimensional hypercube $D=[-1,1]^{2n}$,
then clearly any point not on the unstable manifold will escape in
the $x_{1}$ direction corresponding to the eigenvalue with maximal
real part.  We exploit the fact that the dynamics is governed
by the most unstable direction by assuming $\left|\lambda_{{\rm max}}\right|>>\left|\lambda_{i}\right|$,
$i=2\cdots n$.  If the initial condition lies within a distance $\delta$ of
the unstable manifold with $0<\delta<<1$, then the time to escape from the domain
for an arbitrary non-zero initial condition is given by 
\begin{equation}
t_{f}\approxeq-\frac{\log{(\delta)}}{\lambda_{{\rm max}}}.\label{eq:t_f}
\end{equation}
Using the definition of the exponent given by Eq.~(\ref{eq:FTLE difference defn}),
we have that 
\begin{align}
\sigma(t_{f},{\bm{x}}_{0})=&-\lambda_{{\rm
    max}}\ln{\left(\left|\frac{\epsilon_{1}}{\delta}\right|^{2}+\left|\delta^{-\frac{\lambda_{2}}{\lambda_{{\rm max}}}}\epsilon_{2}\right|^{2}+\cdots+\left|\delta^{-\frac{\lambda_{n}}{\lambda_{{\rm max}}}}\epsilon_{n}\right|^{2}+\left|\delta^{\frac{\lambda_{n}}{\lambda_{{\rm max}}}}\epsilon_{n+1}\right|^{2}\right. } \nonumber\\ 
&\left. +\cdots+\left|\delta^{\frac{\lambda_{2}}{\lambda_{{\rm
            max}}}}\epsilon_{2n-1}\right|^{2}+\left|\delta\epsilon_{2n}\right|^{2}\right)/\left (2\ln{(\delta)}\right ).\label{eq:LLE} 
\end{align}


Since $\left|\lambda_{{\rm max}}\right|>>\left|\lambda_{i}\right|$, and
since $\pm\lambda_{{\rm max}}$ dominates the expanding and contracting directions,  
then for $\delta$ small, we may just consider 
\begin{equation}
\sigma(t_{f},{\bm{x}}_{0})=\frac{-\lambda_{{\rm
max}}\ln{(\epsilon_{1}^{2}/\delta^{2} + \epsilon_{2n}^2\delta^2)}}{2\ln{\delta}}.\label{eq:FTLE
computed}
\end{equation}
Furthermore, we find that
\begin{equation}
\frac{\partial\sigma(t_{f};{\bm{x}}_{0}(\delta))}{\partial\delta}={\frac{\lambda_{{\rm
        max}}\,\ln\left({\epsilon}_{1}^{2}\right)}{2\delta\left(\ln{\delta}\right)^{2}}}\left (1+\frac{\delta^4\epsilon_{2n}^2}{\epsilon_1^2} \right )+\mathcal{O}\left (\frac{\delta^3}{\ln{\delta}}\right ),\
\label{eq:FTLE compyted derivative}
\end{equation}
which can be shown to be negative assuming $\epsilon_{1}<<1$.  Therefore, the FTLE as a function
of distance to the stable invariant manifold is a decreasing function,
and thus takes it maximum values on the manifold.

\section{Conclusions} \label{sec:conc}

In this article, we have considered the dynamics of general stochastic epidemic
models and their extinction properties in finite populations.  The
random fluctuations considered
were from both internal fluctuations, which arise from  mass action
kinetics,  as well as
external random forces, which may be
due to random population  migrations. By examining the extinction processes
from the master equation perspective, eikonal approximations in the large
population limit  give a way to solve for the probability distribution as a
function of time.   A variational principle applied to the exponent of the
  probability distribution near the steady state was used to maximize
the probability to extinction from a disease endemic state.

Maximizing the probability to extinction means minimizing the action, which in
turn generates a Hamiltonian (or Lagrangian) formulation that determines the
flow from endemic to extinct states. The formulation describes the random
fluctuations as a deterministic effective force that overcomes the instability
of the extinct state. Such a deterministic flow describes the
optimal path to extinction from an endemic state. 
Using the above variational formulation,
we explicitly derived the equations of motion describing
the optimal path to extinction in three different models from epidemiology as
a two point boundary value problem

\begin{figure}[t!]
\begin{center}
 \includegraphics[scale=0.25]{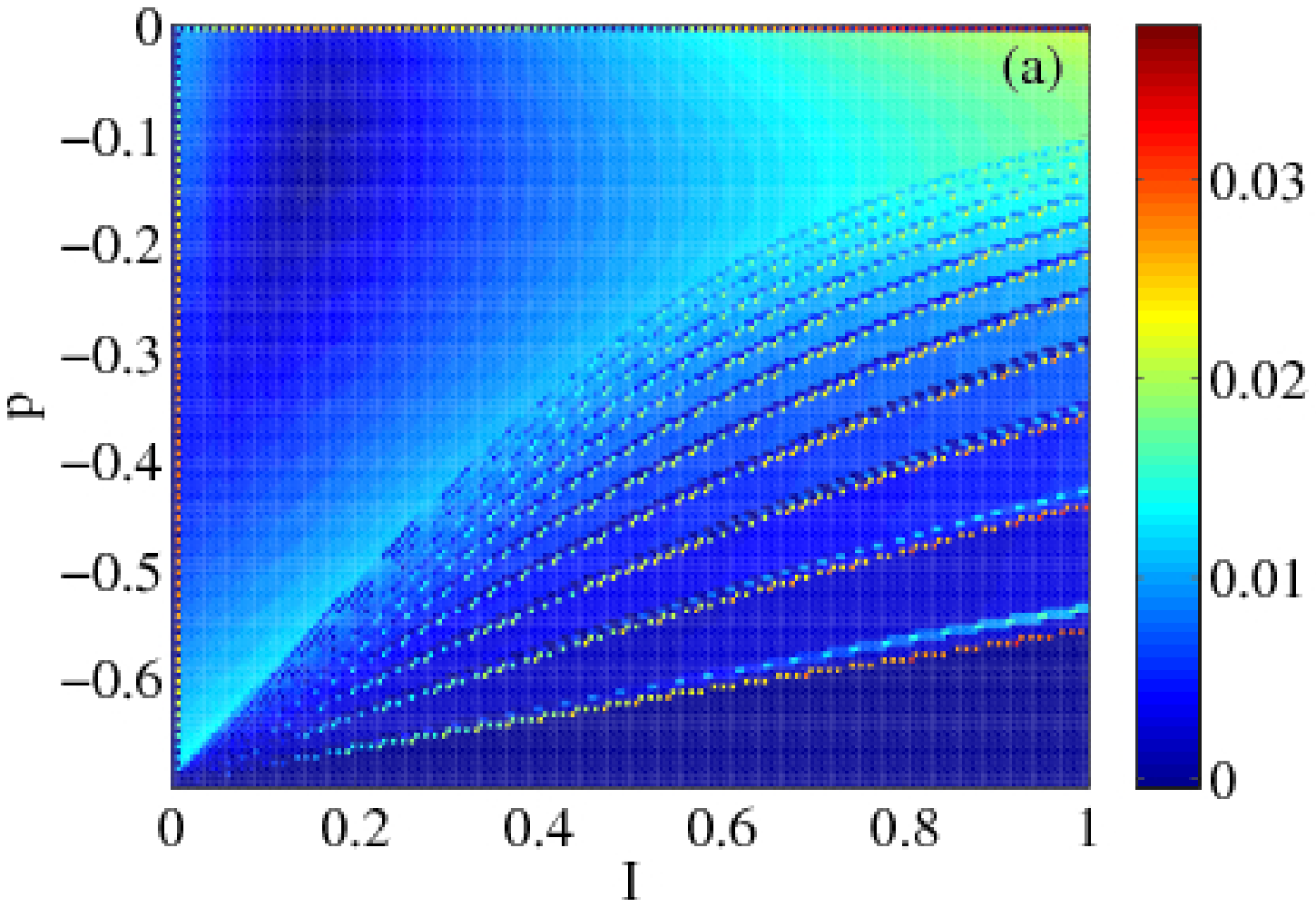}
 \includegraphics[scale=0.25]{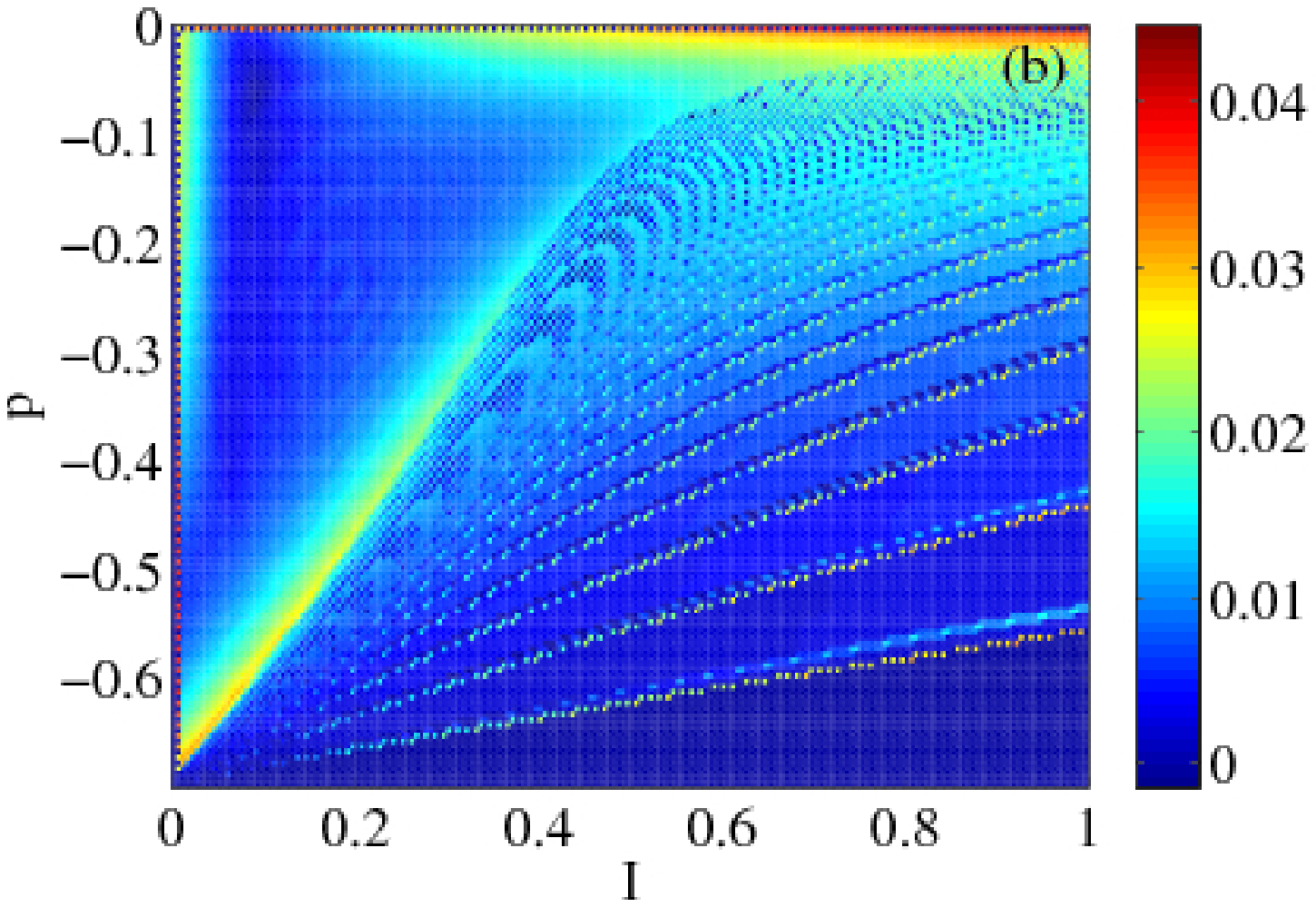}
\includegraphics[scale=0.25]{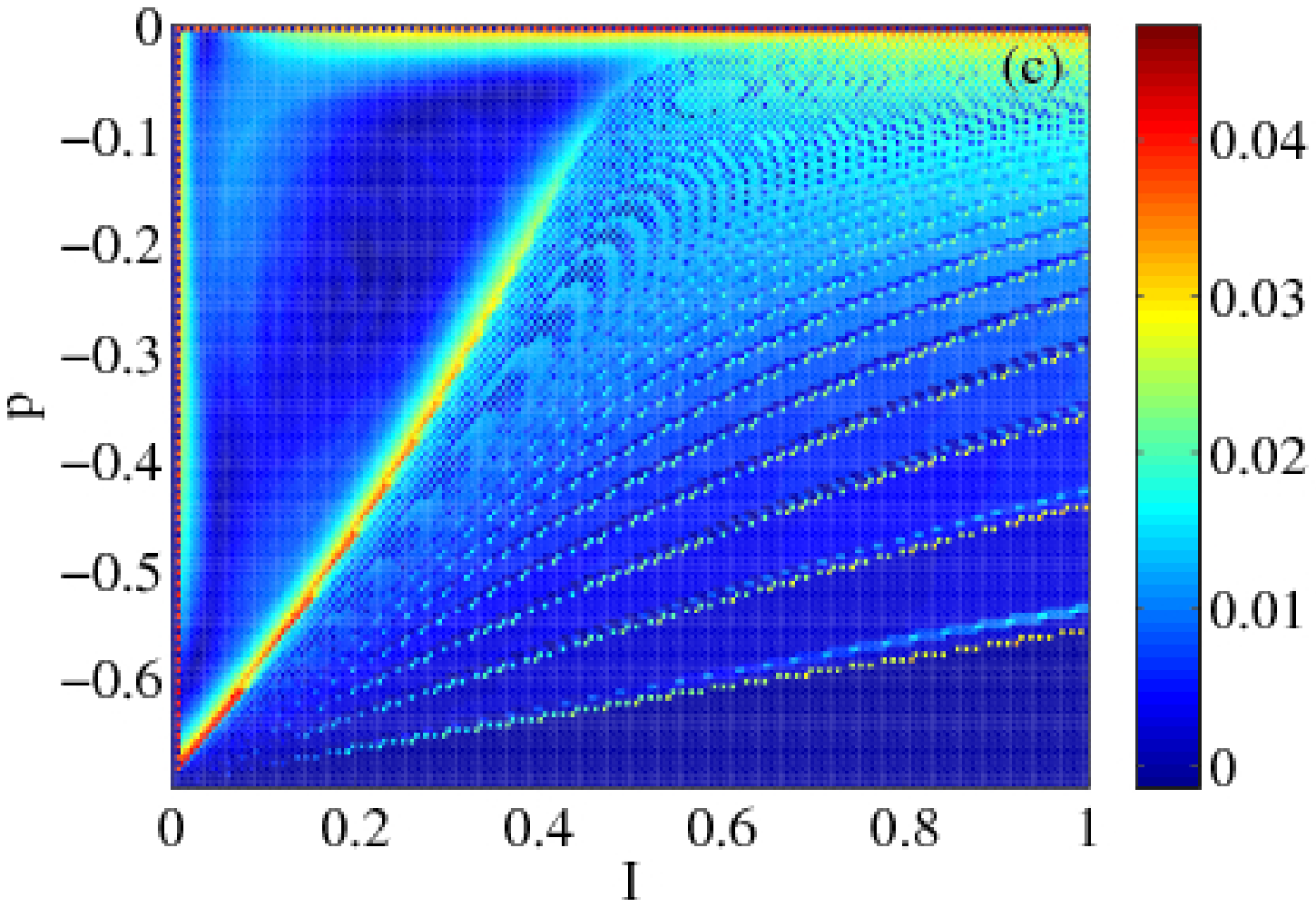}\\
\includegraphics[scale=0.25]{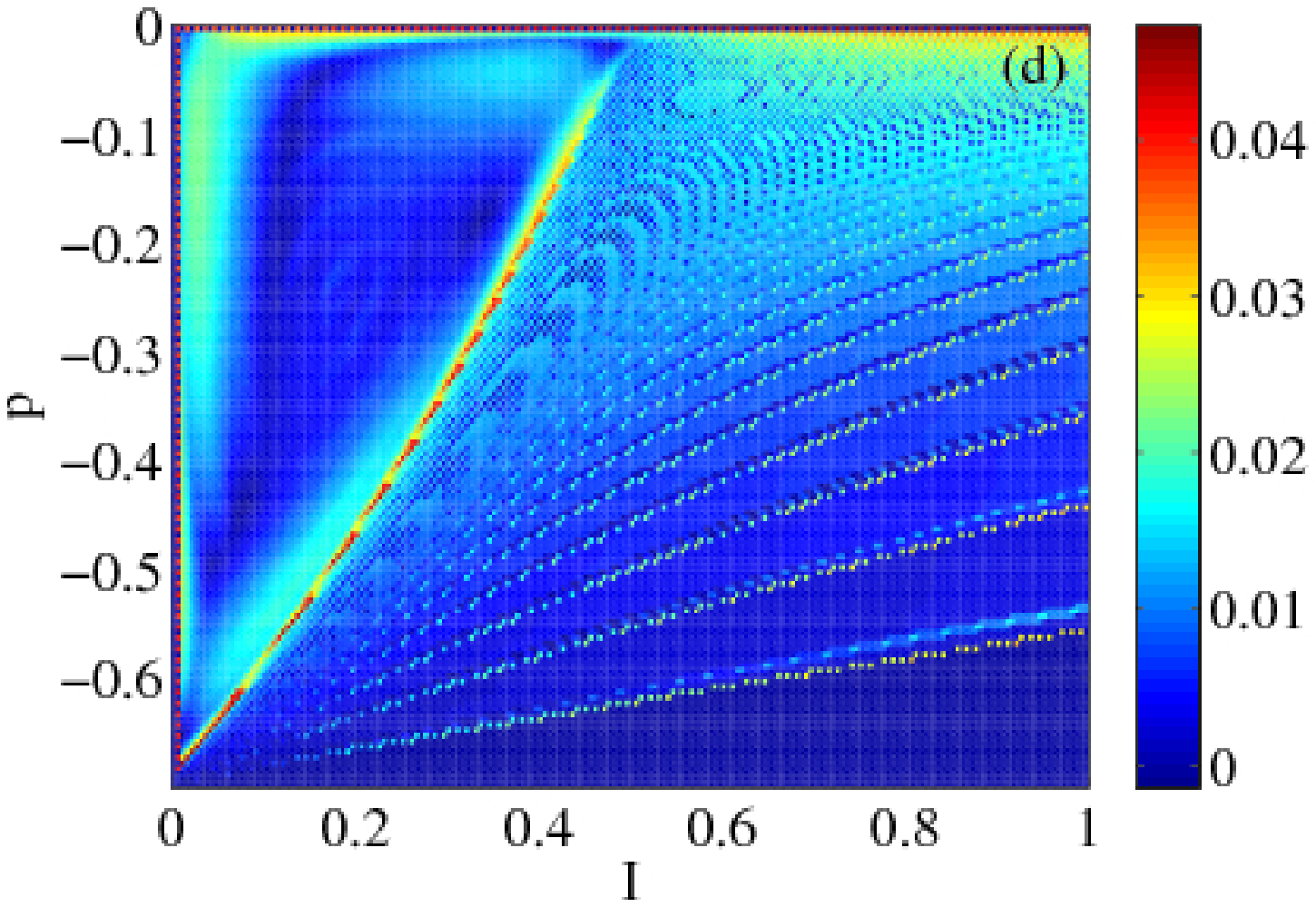}
\includegraphics[scale=0.25]{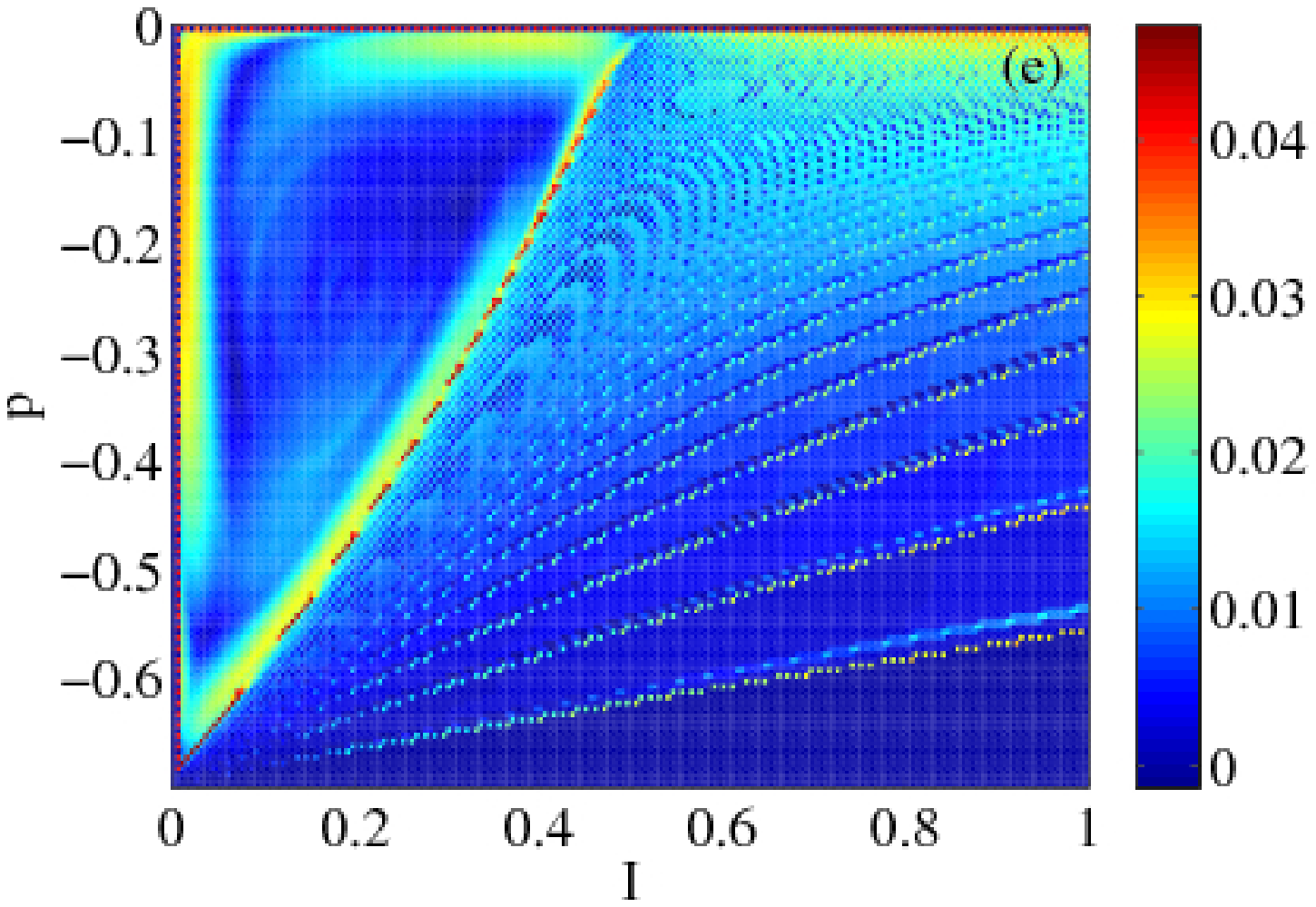}
\caption{\label{fig:FTLE_snaps}(Color online) Snapshots taken at (a) $t=1$, (b)
  $t=3$, (c) $t=5$, (d) $t=7$, and (e) $t=9$ showing the evolution of the FTLE flow field computed using
Eqs.~(\ref{e:Idot_HE})-(\ref{e:pdot_HE}) with $\beta=2.0$ and
$\kappa=1.0$. The integration time is $T=10$ with an integration
step size of $t=0.1$ and a grid resolution of $0.005$ in both $I$
and $p$. The final snapshot in the series is taken at $t=10$ and is shown in
Fig.~\ref{fig:FTLE_Ham}.  A
video that shows the evolution of the FTLE flow field can be found online as
ancillary material.}
\end{center}
\end{figure}

The main result of our paper is that the optimal path is intimately related to
the  maximal sensitivity  of the dynamics of  unstable  Hamiltonian
flows. Specifically,  we introduced a novel method to find the optimal path to
extinction that relies on the computation of the finite-time Lyapunov
exponent field for the dynamical flow  under consideration.  The exponents
provide a measure of sensitivity to initial conditions in finite time.
Moreover, we have shown that the system possesses maximal sensitivity
near the optimal path to extinction.  Therefore, we are able to use
the finite-time Lyapunov exponents to dynamically evolve toward the
optimal path trajectory.

To demonstrate the equivalence of the maximal sensitivity and the optimal path
maximizing the probability of extinction,  we have considered three
prototypical examples from mathematical epidemiology.  In the examples, we
have considered both internal and external noise, and we have
considered both  a 
Hamiltonian and Lagrangian formulation. 
Furthermore, in each of the
three examples, we have shown that the optimal path to extinction
is equated with having a (locally) maximal sensitivity to initial
condition,  which
  implies a  relation  at a fundamental level between the
optimal path and the FTLE.   Even though there exist many possible paths
to extinction, the dynamical systems approach converges to the path
that maximizes the probability to extinction.  

An example of the evolution of the FTLE flow field showing the convergence of
the locally maximal FTLE ridge to the optimal path is shown in
Fig.~\ref{fig:FTLE_snaps}.  The FTLE flow field is computed for $T=10$ using
Eqns.~(\ref{e:Idot_HE})-(\ref{e:pdot_HE}) for the SIS epidemic model with
internal fluctuations (Sec.~\ref{sec:SIS_H}).  Figure~\ref{fig:FTLE_snaps}
shows snapshots of the FTLE field taken at $t=1$, $t=3$, $t=5$, $t=7$, and
$t=9$.  The final snapshot at $t=10$ is shown in Fig.~\ref{fig:FTLE_Ham}.  A
video that shows the evolution of the FTLE flow field can be found online as
ancillary material.  

The parameter values
chosen for the three examples are
such that the extinct and endemic states are far away from  one another, implying
that the system is operating far from any bifurcation points. 
  This result is very important, since in general, no
  approximate analytical treatment, like the one performed in~\citet{dyscla08},  is
  possible if the system's dynamics is not sufficiently close to the
  bifurcation point. 
  However, any scaling behavior of the exponent in the
probability of extinction may still be computed along the optimal
path, which is the important advance afforded by the new
  procedure proposed in this paper.

In the future, we plan on considering more complicated systems.  Of particular
interest are bistable systems (e.g. adaptive networks~\citep{Shaw2008} and the
Schl{\" o}gl birth-death process~\citep{Doering2007}), and higher-dimensional
systems (e.g. multistrain epidemic models~\citep{shbisc07}).  Because the method is general, and unifies dynamical
systems theory with the probability of extinction, we expect that any system found in other fields can be understood using this
approach.

\begin{acknowledgements}
We gratefully acknowledge support from the Office of Naval Research, the Army
Research Office, the Jeffress Memorial Trust, and the
Air Force Office of Scientific Research.  E.F. is supported by a
National Research Council Research Associateship.  L.B.S is supported by Award
Number R01GM090204 from the National Institute Of General Medical
Sciences.  The content is solely the responsibility of the authors and does
 not necessarily represent the official views of the National Institute Of
 General Medical Sciences or the
National Institutes of Health.  We also gratefully acknowledge M. Dykman for
helpful discussions.
\end{acknowledgements}


\end{document}